\documentstyle[12pt,thmsa,a4,sw20lart]{article}

\input{tcilatex}
\begin{document}

\title{Teleportation Scheme of $S$-level Quantum Pure States by Two-level EPRs%
\thanks{%
Supported by National Natural Science Foundation of China under Grant No.
19975043}}
\author{Jindong Zhou and Yong-de Zhang \\
{\it Department of Modern Physics, USTC, Hefei, Anhui 230027, P.R.China}}
\date{(April 25, 2000)}
\maketitle

\begin{abstract}
{\small Unknown quantum pure states of arbitrary but definite }$s${\small %
-level of a particle can be transferred onto a group of remote two-level
particles through two-level EPRs as many as the number of those particles in
this group. We construct such a kind of teleportation, the realization of
which need a nonlocal unitary transformation to the quantum system that is
made up of the s-level particle and all the two-level particles at one end
of the EPRs, and measurements to all the single particles in this system.
The unitary transformation to more than two particles is also written into
the product form of two-body unitary transformations.}
\end{abstract}

Quantum mechanics offers us the capabilities of transferring information
different from the classical case, either for computation or communication.
Bennett {\it et.al. }\cite{bennett}, developed a quantum method of
teleportation, through which, an unknown quantum pure state of a spin-$\frac
12$ particle (we call it 'qubit' \cite{Schumacher1,Lloyd1} ) is teleported
from the sender 'Alice' at the sending terminal onto the qubit at the
receiving terminal where the receiver 'Bob' need to perform a unitary
transformation on his qubit. At first it is necessary to prepare two spin-$%
\frac 12$ particles in an Einstein-Podolsky-Rosen (EPR) entangled state \cite
{Einstein} or so-called a Bell state and send them to the two different
places to establish a quantum channel between Alice and Bob. The second step
is that Alice performs a Bell operator measurement \cite{Braunstein1} to the
quantum system involving her share of the two entangled particles together
with the particle at an unknown state to be transferred. Then through
classical channels, for example, by broadcasting, Alice needs to let Bob
know which one she gets of the four possible outcomes of the Bell operator
measurement. After Bob performs on his share of the two formerly entangled
particles one of four unitary transformations determined by those outcomes,
this particle will be in the unknown state. In this way, the unknown state
is teleported from one place to another.

The new method of teleportation has interested a lot of research groups.
They at once started the research work on quantum teleportation and have
made great development, theoretical and experimental as well. It was
generalized to the case of continuous variables \cite{Vaidman,Kimble}. Sixia
Yu {\it et.al.}, investigated canonical quantum teleportation of
finite-level unknown states by introducing a canonically conjugated pair of
quantum phase and number \cite{Yu1}. The successful experimental realization
of quantum telepartation of unknown polarization states carried on a photon 
\cite{Pan} and the succedent experiments about finite-level quantum system
teleportation \cite{Boschi,Nielsen} have aroused a series of discussions 
\cite{Braunstein2,Pan2,Vaidman2} and further research of this topic from
various aspects \cite{Stenholm,Karlsson,Ralph}. Possible applications have
been considered in Ref. \cite{Zubairy, Maierle}. The method of teleportation
in the case of continuous variables \cite{Vaidman} got its experimental
realization in 1998 \cite{Furasawa}.

From a general point of view, no matter what form it is, there are four
steps to realize the quantum teleportation, which can be seen clearly in
Bennett's initial scheme \cite{bennett}: (a) EPR entangled states preparing;
(b) Bell operator measurements by the sender; (c) the sender informing the
receiver of his outcomes through classical channels; (d) the receiver
performing unitary transformation according to the classical information.
However, the (b) step is not necessary, for it can be substituted by a
nonlocal unitary transformation along with local measurements (Here 'local'
means to single particles). More specifically, the unitary transformation is
performed on the sender's EPR particle and the state-unknown particle to
form some sort of entangled state involving the state-unknown particle
together with the EPR particles, both Alice's and Bob's, while the local
measurements are performed one by one on Alice's particles. These
measurements will result in the random collapse of all the sender's
particles onto definite states. At the other end of communication, the
receiver will got the same results as in the case of performing Bell
operator measurements. In other words, the unitary transformation and local
measurements is equivalent to a Bell operator measurement. The unnecessity
of the (b) step gets further evidence from Ref. \cite{Brassard} in which
Brassard {\it et. al.} indicated the possibility of realizing teleportation
by controlled NOT gates and single qubit operations used in quantum networks.

In this article, it is supposed that the unknown state to be transferred is
an arbitrary but definite $S$-level pure quantum state carried on one
particle labelled with $C$. Different from Ref. \cite{Yu1}, in which the
shared state is a maximally entangled EPR states of $S$-level, we use the
multi-channel made up of $L$ two-level EPRs. It means that at first Alice
and Bob have to prepare this group of EPRs and share each of them, with one
particle of each EPR controllable to the sender and the other to the
receiver. We shall see how the unknown state of $S$-level is teleported from 
$C$ at Alice's place to the Bob halves of the EPRs. It is necessary here to
indicate that the two Hilbert spaces are not the same, one is the single
particle's while the other is the multi-particle's, but from the Hilbert
space with more dimensions (the bigger one) we can always select a subspace
equivalent to the other (the smaller one). In our case, $2^L\geq S$ is
required and therefore we can select $S$ normalized orthogonal vectors as
the basis of the subspace from the $L$ two-level particles' Hilbert space to
make them mapping one by one to the $S$ eigenvectors of $C$. Two states
respectively in the two sorts of Hilbert space will be regarded as the same
if the coefficients are the same when expressed as the linear superposition
of their own basis. Only in this means can we say that the state on $C$ is
teleported onto the $L$ particles.

We label all the EPRs with serial numbers $0,1,\cdots ,L-1$, while the
corresponding particles at Alice's place and Bob's are labelled respectively 
$A_0,A_1,\cdots ,A_{L-1}$ and $B_0,B_1,\cdots ,B_{L-1}$. The EPR entangled
state of each pair of particles $A_k$ and $B_k$ ( $k=0,1,\cdots ,L-1$ ) can
be chosen as follows 
\begin{equation}
\left| \Phi \right\rangle _{A_kB_k}=\frac 1{\sqrt{2}}\left( \left|
0\right\rangle _{A_k}\left| 0\right\rangle _{B_k}+\left| 1\right\rangle
_{A_k}\left| 1\right\rangle _{B_k}\right)  \label{EPR}
\end{equation}
where we express the eigenvectors of the two-level particles as $\left|
0\right\rangle ,\left| 1\right\rangle $ which in the case of $\frac 12$-spin
particles, for example, refer to spin-up state and spin-down state
respectively.. Moreover, the state of $C$ is generally written as 
\begin{equation}
\left| \psi \right\rangle _C=\sum\limits_{m=0}^{S-1}\alpha _m\left|
m\right\rangle _C  \label{stateC}
\end{equation}
in which $\alpha _m$ $(m=0,1,\cdots ,S-1)$ is a complex number satisfying $%
\sum\limits_{m=0}^{S-1}\left| \alpha _m\right| ^2=1$ and $\left|
0\right\rangle ,\left| 1\right\rangle ,\cdots ,\left| S-1\right\rangle $
denote the $S$ eigenvectors of the $S$-level particle. It is convenient that
we distinguish $\left| 0\right\rangle $ and $\left| 1\right\rangle $ only
with subscript, i.e. $\left| 0\right\rangle _{A_k}$ or $\left|
0\right\rangle _{B_k}$ is not the same state with $\left| 0\right\rangle _C$%
, and so is $\left| 1\right\rangle _{A_k}$ or $\left| 1\right\rangle _{B_k}$
with $\left| 1\right\rangle _C$. Further restriction $2^{L-1}<S$ is set on $%
L $, since so many EPRs is the least but enough to realize our teleportation.

Any number can be expressed as its binary form above which we will mark the
symbol '$-$'. For example, a number customarily in decimal form $n$ is
decomposed into $L$-bit number $n=2^{L-1}\cdot n_{L-1}+\cdots +2^1\cdot
n_1+2^0\cdot n_0$ where $2^L\geq n$ and $n_k=0$ or $1$ $(k=0,1,\cdots ,L-1)$%
, and is written as 
\begin{equation}
n=\overline{n_{L-1}\cdots n_1n_0}  \label{binary}
\end{equation}
On the other hand, any binary number has its decimal correspondence. If we
regard the $L$ particles $A_0,A_1,\cdots ,A_{L-1}$ or $B_0,B_1,\cdots
,B_{L-1}$ as 'qubits' \cite{Schumacher1,Lloyd1}, each state $\left|
n_{L-1}\right\rangle _{A_{L-1}}\cdots \left| n_1\right\rangle _{A_1}\left|
n_0\right\rangle _{A_0}=\left| n_{L-1}\cdots n_1n_0\right\rangle _A$ or $%
\left| n_{L-1}\right\rangle _{B_{L-1}}\cdots \left| n_1\right\rangle
_{B_1}\left| n_0\right\rangle _{B_0}=\left| n_{L-1}\cdots
n_1n_0\right\rangle _B$ ($n_k=0$ or $1$, $k=0,1,\cdots L-1$) will correspond
to a binary number $\overline{n_{L-1}\cdots n_1n_0}$ and we introduce a
symbol '$\left| \text{ }\right\rangle \rangle $' to simplify the denotation
of the state as 
\begin{equation}
\left| n\right\rangle \rangle \equiv \left| n_{L-1}\cdots n_1n_0\right\rangle
\label{simden}
\end{equation}
where $n$ has the same meaning as in Eq. \ref{binary}. The quantum state of
the composite system made up of $A$, $B$ and $C$ will thus be as follows 
\begin{equation}
\left| \Psi _0\right\rangle _{ABC}=\left| \psi \right\rangle
_C\prod\limits_{k=0}^{L-1}\left| \Phi \right\rangle _{A_kB_k}=\frac 1{\sqrt{N%
}}\sum\limits_{m=0}^{S-1}\sum\limits_{n=0}^{N-1}\alpha _m\left|
m\right\rangle _C\left| n\right\rangle \rangle _A\left| n\right\rangle
\rangle _B  \label{ABC}
\end{equation}
where $N=2^L$.

In principle, Alice is able to perform on the composite system $AC$ any
quantum operations, including local or nonlocal unitary transformations and
measurements. To realize the teleportation, a nonlocal unitary
transformation $U_{AC}$ to all the bodies included in system $AC$ is
performed. $U_{AC}$ will realize the following transformation 
\begin{equation}
U_{AC}\left| m\right\rangle _C\left| n\right\rangle \rangle _A=\frac 1{\sqrt{%
S}}\sum_{j=0}^{S-1}e^{i\frac{2mj\pi }S}\left| j\right\rangle _C\mid
f^n(j,m)\left\rangle {}\right\rangle _A  \label{UT}
\end{equation}
in which $m=0,1,\cdots ,S-1$, and $f^n(j,m)$ is a number of decimal form
determined by $j$, $m$ and $n$ so that $\mid f^n(j,m)\left\rangle
{}\right\rangle $ is one of the $N$ eigenstates. If we also express $j$, $m$
and $f^n(j,m)$ as the binary form 
\begin{eqnarray}
j &\equiv &\overline{j_{L-1}\cdots j_1j}_0  \label{binary2} \\
m &\equiv &\overline{m_{L-1}\cdots m_1m_0}  \nonumber \\
f^n(j,m) &\equiv &\overline{f_{L-1}^n(j,m)\cdots f_1^n(j,m)f_0^n(j,m)} 
\nonumber \\
j_k,m_k,f_k^n(j,m) &=&0,1(k=0,1,\cdots .L-1)  \nonumber
\end{eqnarray}
$f^n(j,m)$ will be determined by $f_k^n(j,m)$s that satisfy 
\begin{equation}
f_k^n(j,m)=n_k\oplus j_k\oplus m_k  \label{module}
\end{equation}
where '$\oplus $' denotes addition modulo $2$. One can easily prove the
unitarity of $U_{AC}$ and show that when any two among $j$, $m$ and $n$ are
definite, $\mid f^n(j,m)\left\rangle {}\right\rangle $s different in the
parameter of the rest will be orthogonal mutually. For example, 
\begin{equation}
\left\langle {}\right\langle f^n(j,m^{\prime })\mid f^n(j,m)\left\rangle
{}\right\rangle =\delta _{m^{\prime }m}  \label{ortho}
\end{equation}
Where $m,m^{\prime }=0,1,\cdots ,S-1$. Eq. \ref{ortho} means that any two
basis among $\mid f^n(j,m)\left\rangle {}\right\rangle $s with the same $n$
and $j$ but different $m$ will not be the same.

After the transformation of $U_{AC}$, due to Eq. \ref{UT}---\ref{module},
the quantum state of system $ABC$ will change to 
\begin{equation}
\left| \Psi \right\rangle _{ABC}=U_{AC}\left| \Psi _0\right\rangle
_{ABC}=\frac 1{\sqrt{S}}\sum_{j=0}^{S-1}\left\{ \left| j\right\rangle
_C\frac 1{\sqrt{N}}\sum_{n=1}^{N-1}\left( \left| n\right\rangle \rangle
_A\sum_{m=0}^{S-1}\alpha _me^{i\frac{2mj\pi }S}\mid f^n(j,m)\left\rangle
{}\right\rangle _B\right) \right\}  \label{stateAT}
\end{equation}
which is the entangled quantum state involving all the particles in system $%
ABC$.

If now Alice performs measurements to the single particles $C,A_0,A_1,\cdots
,A_{L-1}$, with the same possibility of $\frac 1{NS}$, she will acquire one
of the outcomes, i.e., the collapse of the state of these particles to the
possible eigenstate $\left| j\right\rangle _C\left| n\right\rangle \rangle
_A $ ($j=0,1,\cdots ,L-1$ and $n=0,1,\cdots ,N-1$). Thus the entanglement
among $A$, $B$ and $C$ will be destroyed and Bob will acquire the state of $%
B $%
\begin{equation}
\left| \psi ^n(j)\right\rangle _B=\sum_{m=0}^{S-1}\alpha _me^{i\frac{2mj\pi }%
S}\mid f^n(j,m)\left\rangle {}\right\rangle _B  \label{Bstate}
\end{equation}
which is an entangled quantum state of particles $B_0,B_1,\cdots ,B_{L-1}$.
If $n$ and $j$ are definite, Eq. \ref{ortho} ensure that we can redefine $S$
of $B$'s basis as $\left| m\right\rangle ^{\prime }\equiv e^{i\frac{2mj\pi }%
S}\mid f^n(j,m)\left\rangle {}\right\rangle $ $\left| m\right\rangle
^{\prime }=e^{i\frac{2mj\pi }S}\mid f^n(j,m)\left\rangle {}\right\rangle $,
where $\left| 0\right\rangle ^{\prime },\left| 1\right\rangle ^{\prime
},\cdots ,\left| S-1\right\rangle ^{\prime }$ form the basis of the subspace
of system $B$'s Hilbert space. Therefore we get 
\begin{equation}
\left| \psi ^n(j)\right\rangle _B=\sum_{m=0}^{S-1}\alpha _m\left|
m\right\rangle _B^{\prime }  \label{redefi}
\end{equation}
According to our discussion in paragraph 4 and the comparison of Eq. \ref
{stateC} and \ref{redefi}, we can regard $\ \left| \psi ^n(j)\right\rangle $
and $\left| \psi \right\rangle $ as the same. However, we need indicate that 
$\left| m\right\rangle ^{\prime }$ lies on $j$ and $n$, which makes it is
still necessary to build the classical channels between Alice and Bob to
transfer the information about Alice's outcomes, or the information of $j$
and $n$ in the other words, since Bob will not know exactly what the $\left|
m\right\rangle ^{\prime }$ means without the knowledge of $j$ and $n$. Just
the necessity of classical information transferring makes the
faster-than-light communication impossible.

We have discussed above the possibility, in principle, the possibility of
teleportation of any $S$-level quantum states by no less than $L=\log S$
two-level EPRs. In our discussion, we use the complicated unitary
transformation $U_{AC}$, which means the evolution of the quantum state of
system $AC$ under the interaction of all those particles involved in $AC$.
The complication of $U_{AC}$ leads to the complication of operation. It is
even impossible for us to operate such a transformation unless we take
further consideration. The method of quantum computational networks has
shown out the most feasible way of realizing the operation. The quantum
computational networks has been much studied in Ref. \cite{Deu1,Deu2,DiV}.
Following their method, we make the transformation more operationable by
decomposing $U_{AC}$, which is to $2L+1$ particles, into a sequence of
two-body unitary transformations and a simple single-body unitary
transformation. Only two classes of such transformations are used: (a) the
discrete Fourier transform modulo $S$, denoted $DFT_S$, which is a unitary
transformation in $S$ dimensions.. It is defined relative to the basis $%
\left| 0\right\rangle _C,\left| 1\right\rangle _C,\cdots ,\left|
S-1\right\rangle _C$ by 
\begin{mathletters}
\begin{equation}
DFT_S\left| m\right\rangle _C=\frac 1{\sqrt{S}}\sum_{j=0}^{S-1}e^{i\frac{%
2mj\pi }S}\left| j\right\rangle _C  \label{DFT}
\end{equation}
(b) a combined unitary transformation $U_{Ck}$ to the two particles $C$ and $%
A_k$ ($k=0,1,\cdots ,L-1$). $U_{Ck}$ is defined by 
\end{mathletters}
\begin{equation}
U_{Ck}\left| m\right\rangle _C\left| n_k\right\rangle _{A_k}=\left|
m\right\rangle _C\left| m_k\oplus n_k\right\rangle _{A_k}  \label{U2}
\end{equation}
$U_{AC}$ can be decomposed into the product of these two classes of
transformation 
\begin{equation}
U_{AC}=\left( \prod_{k=0}^{L-1}U_{Ck}\right) \cdot DFT_S\cdot \left(
\prod_{k=0}^{L-1}U_{Ck}\right)  \label{decompose}
\end{equation}
where because $\left[ U_{Ck^{\prime }},U_{Ck}\right] =0$ for any $%
k,k^{\prime }=0,1,\cdots ,L-1$, we need not distinguish their order. By \ref
{decompose} we simplify the problem in operation of $U_{AC}$, for the
quantum operation on two bodies is far more feasible than on a lot of bodies.

In summary, we construct the scheme of transferring an arbitrary $S$-level
quantum state by using two-level EPRs. The importance of this construction
lies not only on the scheme itself, but also on the possibility of further
research and application of teleportation. It leads us to more general, more
feasible and simultaneously more challenging considerations on the problem
of teleportation. A lot of questions, such as probabilistic teleportation
and teleportation of unknown quantum states by definite number of EPRs, are
thus put forward before us, waiting for us to solving.

\begin{center}
{\Large ACKNOWLEDGMENTS}
\end{center}

We would like to thank all the other members of our research group for their
helpful discussions. They are Guang Hou, Shengjun Wu, Prof. Qiang Wu, Yifan
Luo, Minxin Huang, Miss Jie Yang, Guojun Zhu, Ganjun Zhu and Meisheng Zhao.
We have also benefited from the cooperation with Prof. Anton Zeilinger's
group at Austria. We are grateful to them especially Dr. Jianwei Pan for
their helpful information and suggestions.

\end{document}